# Electronic transition from graphite to graphene via controlled movement of the top layer with scanning tunneling microscopy

P. Xu,[1] Yurong Yang,[1,2] D. Qi,[1] S.D. Barber,[1] J.K. Schoelz,[1] M.L. Ackerman,[1] L. Bellaiche,[1] and P.M. Thibado[1,*]

[1]Department of Physics, University of Arkansas, Fayetteville, Arkansas 72701, USA

[2]Physics Department, Nanjing University of Aeronautics and Astronautics, Nanjing 210016, China

A series of measurements using a technique called electrostatic-manipulation scanning tunneling microscopy (EM-STM) were performed on a highly-oriented pyrolytic graphite surface. The electrostatic interaction between the STM tip and the sample can be tuned to produce both reversible and irreversible large-scale movement of the graphite surface. Under this influence, atomic-resolution STM images reveal that a continuous electronic transition from triangular symmetry, where only alternate atoms are imaged, to hexagonal symmetry can be systematically controlled. Density functional theory (DFT) calculations reveal that this transition can be related to vertical displacements of the top layer of graphite relative to the bulk. Evidence for horizontal shifts in the top layer of graphite is also presented. Excellent agreement is found between experimental STM images and those simulated using DFT.

# I. INTRODUCTION

When a material is cut to form a surface, the broken bonds tend to rearrange into a lower energy configuration. This process is known as a surface reconstruction and results in the surface atoms having a different symmetry from the bulk atoms. For example, on the Si(001) surface adjacent Si atoms will tilt toward each other to form a dimer bond. In doing so, half of the broken surface bonds can be reformed to significantly lower the total surface energy. The symmetry of the surface is now different from the bulk, since the periodicity along the dimer bond is twice the bulk, thereby yielding a (2×1) surface reconstruction.[1] Similar things happen on the GaAs(001) surface,[2] but here the atomic arrangement is dependent on the arsenic pressure as well as the substrate temperature. In some instances a phase transition can be identified between the various reconstructions.[3] At the other extreme, a more subtle surface reconstruction can occur which involves only the electron distribution within the material. A prime example is the easily cleaved GaAs(110) surface, which exhibits very weak bonding between layers.[4] Because of this, when the layers are separated, the atomic nuclear positions remain the same, but the surface charge density significantly redistributes itself. The charge shifts to be only on the surface As atoms instead of the being equally shared between the Ga and As atoms. Consequently, scanning tunneling microscopy (STM) filled-state images show only the As atoms while empty-state images show only the Ga atoms.

Similar to GaAs, highly-oriented pyrolitic graphite (HOPG) is another example of a system which is easily cleaved. When HOPG is imaged using STM, only alternate atoms contribute to the tunneling current. This results in an image with triangular symmetry rather than the expected hexagonal symmetry. This surprising result is attributed to the particular stacking order most commonly observed in hexagonal graphite,[5] referred to as ABA or Bernal stacking.

Half of the surface carbon atoms (the A atoms) are directly above atoms in the layer below, while the other half (the B atoms) are directly above hexagonal holes. The electronic charge density of the A atom is pulled into the bulk, and consequently the STM is unable to image it.[6] However, when a single layer of graphite is separated from the bulk, the symmetry is restored, and the subsequent redistribution of the electron density allows every carbon atom to be imaged with the STM. This real-space transformation also leads to all the well-known electronic properties which distinguish graphene from graphite,[7] including a band structure with linear rather than parabolic dispersion.[8]

Transitions to a linear band structure are especially interesting because the charge carriers lose their effective mass. This is a process of fundamental importance in physics. Something similar to this transition has been observed in bilayer graphene using electrical transport measurements. Lau and coworkers recently demonstrated that bilayer graphene undergoes a phase transition at a critical temperature of 5 K to an insulating state with a band gap of ~3 meV.[9] The effect can be tuned or reversed with the application of a perpendicular electric field or magnetic field.[10]

Studies using graphite have observed similar things; however, the events are randomly occurring. For example, using low-temperature (4.4 K) STM, low-voltage scanning tunneling spectroscopy (STS), and a magnetic field, Landau levels consistent with graphene have been observed on graphite by Andrei and coworkers.[11,12] Signatures in the sequence have been used to quantitatively predict the amount of interaction between the top layer and the bulk. Further evidence of varying degrees of coupling can be seen in the symmetry of STM images. The STM tip can provide a perturbation that vertically lifts the top layer, resulting in images which exhibit a range of possibilities between the triangular and hexagonal lattices.[13] The difficulty, however, is that this induced decoupling has been random, not lending itself to a systematic study of the important symmetry-breaking transition from bulk graphite to monolayer graphene.

A surface charge density similar to graphene but on graphite can also be attributed to horizontal shifts in the surface layer.[14,15] This has created a lot of excitement in potentially controlling the stacking of graphene layers. For example, recent work suggests that stacking graphene is a way to solve the band gap problem, which is currently the chief obstacle for using graphene in digital electronic devices. Trilayer graphene is especially interesting because two stable allotropes have been identified; the layers can be arranged with ABA (Bernal) stacking or ABC (rhombohedral) stacking. ABC trilayers exhibit an inherent band gap of ~6 meV at the K-point, which can be increased by applying an electric field, while no such band gap is predicted in ABA trilayers.

Naturally, several major steps have been taken toward characterizing the stacking sequence. For instance, Raman spectroscopy performed on mechanically exfoliated graphene has revealed that the majority of the trilayers produced are ABA stacked, while only about 15% are in the ABC configuration.[16,17] On the other hand, when graphene is grown on SiC (0001), the layers selectively form in the ABC order over ABA, as observed with high-resolution transmission electron microscopy.[18] Certainly, one would like to control the stacking sequence or ideally alter it from one form to the other. A related area with a lot of interest is rotated or twisted layers.[19,20] This has a lot of appeal because all the physics can be parameterized with just one angle. Horizontal shifting has received less attention.[21]

In this article, we present STM images of the HOPG surface before, during, and after perturbing the surface using a technique called electrostatic-manipulation STM (EM-STM).[22] With this technique large-scale precision-controlled vertical movement of the graphite surface is possible. Atomic-scale STM images reveal a continuous transition from graphite to graphene. Density functional theory (DFT) calculations were used to generate a complete set of simulated STM images and provide excellent agreement with the measurements. The continuous change in the spatial distribution of the charge density is proposed as a measure of coupling between the surface layer and bulk. Next, we present STM images on HOPG surface which show clear evidence of the top layer shifting horizontally in a direction along the carbon-carbon (C-C) bond axis. Excellent agreement with a series of DFT simulated STM images generated from structures shifted along this same direction is presented. From DFT we also find the direction for the lowest-energy barrier to transition from ABA to ABC stacking.

## II. EXPERIMENTS

The experimental STM images and EM-STM line profiles were obtained using an Omicron ultrahigh-vacuum (base pressure is $10^{-10}$ mbar), low-temperature STM operated at room temperature. The top layers of a 6 mm × 12 mm × 2 mm thick piece of HOPG were exfoliated with tape to expose a fresh surface. The HOPG was then mounted with silver paint onto a flat tantalum STM sample plate and transferred into the STM chamber, where it was electrically grounded. STM tips were electrochemically etched from 0.25 mm diameter tungsten wire via a custom double lamella setup using an automatic gravity cutoff circuit.[23] After etching, the tips were gently rinsed with distilled water and dipped into a concentrated hydrofluoric acid solution to remove surface oxides[24] before being transferred into the STM chamber through a load lock. Numerous filled-state STM images of the HOPG surface were acquired using a tip bias of +0.100 V and a constant current of 0.20 nA for small scale images and 1.00 nA for large scale images.

A collection of illustrative graphite STM images is displayed in Fig. 1. All have had minimal image processing are shown with the fast scan direction horizontal and with the slow scan direction going from bottom to top. The last three images have been "flattened" so that each line of data has the same average height. An STM image of the graphite surface measuring 100 nm × 100 nm and with a monolayer step running diagonally across the surface is shown in Fig. 1(a). An atomic-resolution STM image measuring 6 nm × 6 nm showing the traditional triangular symmetry lattice structure for graphite is shown in Fig. 1(b). This is the typical STM image for graphite and is relatively easy to obtain because only every other carbon atom is detected. Sometimes while imaging graphite, all of the carbon atoms can be observed, resulting in the hexagonal symmetry or honeycomb pattern as shown in Fig. 1(c). This is the same pattern one would observe when imaging isolated graphene. Less common are STM images which show both triangular and hexagonal patterns within a single image as shown in Fig. 1(d). The beginning of this scan shows the traditional graphite structure, while after about two-thirds of the scan the surface abruptly switched to the hexagonal structure. Note, the subsequent STM image acquired (not shown) was similar to graphene throughout. An even more interesting result is shown in Fig. 1(e). Here the triangular pattern occurs on the left side, while the hexagonal pattern occurs on the right side. Note, this transformation occurs along the fast scan direction, line-by-line throughout the image acquisition process. Finally, the surface can also start out with a hexagonal pattern and switch to the triangular pattern as shown in Fig. 1(f).

The above STM images have been observed numerous times, over a long time period, and with a plethora of STM tips. They are somewhat randomly occurring; however, the frequency can be increased by increasing the setpoint current or reducing the bias voltage. We believe the properties of the STM tip are not changing throughout these scans, but that the local properties of the graphite sample are changing. For example, line-by-line one can see a clear triangular atomic-resolution pattern along the left edge of Fig. 1(e). Each atom gets rescanned about 10 times (i.e., 400 data points per line with 400 lines per image) to piece together the image of a single atom, and each row of atoms appears in the proper triangular position relative to the next row going up the scan. Simultaneously, and also line-by-line, a clear hexagonal pattern is being observed on the right edge of Fig. 1(e). It is not possible to associate these changes with changes at the end of the STM tip.

Ideally, we would like to have these alterations occurring on the surface at will *versus* at random. We found that performing high-voltage, constant-current scanning tunneling spectroscopy (CC-STS) measurements can lift the top layer of graphite. During a CC-STS measurement, the imaging scanner is paused, and the feedback loop controlling the vertical motion of the STM tip remains operational. As the STM tip bias is varied, one records the vertical displacements required to maintain a constant tunneling current. Assuming the sample is stationary, this process indirectly probes its density of states (DOS).[25] A second interaction is also taking place, though, in which the tip bias induces an image charge in the grounded sample, resulting in an electrostatic attraction that increases with the bias. We have found that in some materials, such as graphite[26] and freestanding graphene,[27] this attraction can result in physical movement of the sample, convoluting and often eclipsing any DOS measurement. Given that it is not a spectroscopy measurement, we call this measurement EM-STM. In an EM-STM experiment, the deformations are actually the subject of interest. By employing electrostatic forces created by the STM tip, one may physically manipulate the surface plane of atoms and examine some of its mechanical properties. Thus, an EM-STM measurement involves recording the z-position of the tip as the bias voltage is varied at constant current, with the goal of controlled sample manipulation.

A diagram with a typical EM-STM data set and some illustrations of how this technique might appear on an atomic scale is shown in Fig. 2. The EM-STM measurement was taken on graphite and is shown in Fig. 2(a). Simultaneously, the tunneling current was also measured, and the result is plotted in the inset diagram. Measuring the tunneling current is critical to proving that the current remains at an approximately constant value of 0.2 nA throughout the duration of the measurement. The EM-STM data shows that during the voltage sweep from 0.1 V to 0.6 V the tip is held at its initial height with little variation. An illustration showing the relative tip-sample position during this period is shown in Fig. 2(b). From 0.6 V to 0.8 V, the tip swiftly retracted by about 30 nm, at which height it roughly stabilized. Notice the tunneling current remains essentially constant, indicating that the sample follows the tip. The sudden movement of the tip suggests that the top layer of graphite is being held in place by the substrate until the electrostatic force of attraction, which increases with voltage, becomes large enough to suddenly separate the layers. A schematic of the initial release due to the STM tip is illustrated in Fig. 2(c). When the voltage rises above 0.8 V, the top graphite layer is significantly lifted by the tip and fully decoupled from the bulk locally as illustrated in Fig. 2(d). Again, the measured tunneling current serves as evidence that the sample surface must move with the tip. If it did not, the current would exponentially fall to zero around 0.6 V. Note, traditional constant-height (feedback off) STS data was also acquired (not shown), but the current quickly saturated the preamplifier at these higher voltages, consistent with the sample crashing into the stationary STM tip.

The EM-STM technique significantly broadens the abilities of the STM, which is already known for its superior ability to obtain atomic structural and local electronic information for rigid samples. Now, if the sample is free to move or suspended, one can use EM-STM to gain insight into the local electrostatic and elastic properties. This could prove valuable when considering chemically modified freestanding graphene, for example.

## III. CONTROLLED VERTICAL MOVEMENT AND ELECTRONIC TRANSITION FROM GRAPHITE TO GRAPHENE

### A. EM-STM on graphite strip

The ability to physically alter the HOPG surface using EM-STM is demonstrated in Fig. 3. A series of 150 nm × 150 nm STM images all at the same location were taken before, during, and after EM-STM measurements, and the images are displayed in sequential order in Fig. 3(a-e). As before, the slow scan direction of the STM tip proceeded from bottom to top, and the images are colored such that the highest points are white (~2 nm high) while the lowest points are black. A white strip approximately 20 nm wide is prominent in Fig. 3(a), indicating that a raised ribbon-like structure exists on the HOPG surface. This image was taken prior to any EM-STM measurements. A darker strip, or trench, can also be seen approximately 50 nm to the right of the white strip, with a protrusion in the trench serving as a reference point when comparing the images. During the next scan, which is presented in Fig. 3(b), an EM-STM measurement was carried out shortly after the scan started. The STM tip was first positioned on the white strip just above the in-progress scanning position, and then the tip bias was increased from 0.1 V to 10.0 V at a constant tunneling current of 1.00 nA. Once the scan resumed, the white strip was found to be displaced to the right, toward the protrusion. Surprisingly, after scanning further the white strip suddenly jumped back to its original position. In the next scan, we found that the lower portion of the white strip had been displaced to the right as shown in Fig. 3(c). Now, however, this section is somewhat darker (it is likely a fold in the ribbon), indicating that a permanent change has been introduced to the surface. To demonstrate this ability again, a second EM-STM measurement was taken during the subsequent scan shown in Fig. 3(d). This resulted in a displacement of the upper portion of the white strip, this time away from the trench. The next scan, taken immediately afterward and shown in Fig. 3(e), shows that a larger portion of the white strip is farther away from the trench, resulting in a structure clearly distinct from that in Fig. 3(a). A larger scale image of the same location further reveals that a permanent change was made to the local region of the surface, as shown in Fig. 3(f). This sequence of images helps illustrate the size of the area that can be impacted by an EM-STM measurement on graphite. The height of the STM tip versus bias voltage, acquired during one of the two EM-STM measurements is shown in Fig. 3(g), with an inset showing that the tunneling current remains roughly constant at 1 nA. The EM-STM measurement shows a continuous increase in the height of the STM tip with increase in bias voltage. In order to determine the approximate electrostatic force between the STM tip and the graphite surface as a function of bias voltage, we modeled the tip and sample using the method of images.[28] The tip is modeled as a biased conducting sphere of radius 20 nm, and the graphite is modeled as an infinite grounded conducting plane. The initial sphere-plane separation was set at 0.5 nm, but this value was adjusted as the voltage increased to correct for the small vertical movement observed in a stationary control sample of graphene on copper foil. The calculated force *vs.* voltage data was then combined with the experimental EM-STM data in Fig. 3(g) to produce the attractive electrostatic force as a function of tip height shown in Fig. 3(h). The force increases almost linearly with bias voltage to a maximum of about 4 nN. The area under the force curve yields an energy cost of about 230 eV being required to move the graphite strip.

**B. EM-STM on pristine graphite terrace**

A series of EM-STM measurements taken on a pristine graphite terrace is shown in Fig. 4(a-e). This set of data shows the range of results that can occur. Typically, the height changes by a small amount when EM-STM first applied to a given location [as shown in Fig. 4(c)]. Subsequent measurements in the same location can cause larger movement to occur [as shown in Fig. 4(e)]. In addition, reversal in the movement can sometimes happen [as shown in Fig. 4(d)]. The calculated electrostatic force as a function of applied bias was used to convert the EM-STM data shown in Fig. 4(a-e) into the force versus height curves shown in Fig. 4(f-j), respectively. The same maximum force of about 0.4 nN is reached in each data set because the same voltage range was used for each data set. Similar to before, the energy expended by the STM can be found from the area under each curve.

**C. Vertical displacement of top layer of graphite: Experiment**

A series of high-magnification, atomic-resolution STM images of the HOPG surface are presented in Fig. 5(a-e). Going down the page, each image shows the gradual transformation from full-triangular to full-hexagonal symmetry. We start with a typical STM image of HOPG as shown in Fig. 5(a). The bright white circles are arranged with perfect triangular symmetry and represent the $p_z$-orbitals of the B-type carbon atoms which are above the hexagonal holes in the second layer. For this image, the unit cell depicts only one atom. Next, a very weakly visible A-type atom can be seen in Fig. 5(b). An asymmetrical hexagonal pattern is starting to appear in Fig. 5(c). Two atoms are now apparent in the unit cell but with a much larger charge density on one atom. A more balanced hexagonal pattern is observed in Fig. 5(d). The nearly perfect honeycomb structure is shown in Fig. 5(e). Here both atoms in the unit cell possess nearly equal charge density, resembling a typical STM image of graphene rather than graphite. This type of image on HOPG is much less common than the triangular one, and in the past obtaining it has been mostly a matter of chance. However, EM-STM provides a mechanism for directly separating the surface layer from the bulk at will, effectively creating a section of graphene. By systematically repeating the EM-STM measurement at successively higher voltages, one can tune the displacement of the top layer. While this procedure does lift the layer, the top layer is still attracted to the graphite and thus quickly relaxes. Nevertheless, the likelihood of observing the graphene hexagonal symmetry on graphite does greatly increase after repeatedly performing EM-STM.

### D. Vertical displacement of top layer of graphite: Simulation

A full understanding of our experimental findings shown in Fig. 5(a-e) was not possible until simulated STM images of HOPG were extracted from DFT calculations.[29] These calculations were performed within the local-density approximation without modeling the STM tip[30] and using projector augmented-wave potentials[31] as implemented in the plane wave basis set VASP[32] code. The graphite was modeled as a six-layer Bernal stack, using a $1 \times 1$ unit cell. A cutoff energy of 500 eV and a very large $219 \times 219 \times 1$ Monkhorst-Park k-point mesh were used to ensure proper sampling around the Dirac point. Initially, the atoms were allowed to move until all forces were less than 0.1 eV/nm, resulting in a C-C bond length of 0.142 nm and an interplanar separation of 0.334 nm. Then, the top layer was moved away from the bulk in ten equal steps of size 0.015 nm, allowing only in-plane relaxation at each step. For each configuration, a simulated constant-current STM image was produced by integrating the local

DOS from the Fermi level to 0.06 eV below that point and choosing an appropriate isocontour surface. These parameters were chosen to best replicate the experimental STM conditions.

Six simulated STM images taken from the DFT calculations are presented in Fig. 5(f-k). For each, the displacement of the top plane relative to its equilibrium position is noted. Large spheres representing the electron density around the B-type atoms arranged in a triangular pattern are shown in Fig. 5(f). This is known to be due to half of the top layer carbon atoms (A-type) forming a dimer bond with the carbon atom directly underneath it in the plane below. This bonding is responsible for reducing the surface charge density of these atoms. The much smaller and less bright adjacent features (shaped like a small triangle) represent the electron density around the A-type atoms, but this is not resolved in the experimental STM images. For net vertical displacements including 0.015 nm up through 0.045 nm, the circles shrink while the triangles grow larger and more rounded. As we continue down the column to the 0.135 nm displacement, we see that the electron density for each atom becomes essentially equivalent. Further displacements do not result in any additional changes. The simulated images are in excellent agreement with the experimental data.

The side views of the six-layer simulated structure with the top layer moving from a net displacement of 0.00 nm to 0.135 nm are shown in Fig. 5(l-q). Notice how the charge density of the top layer becomes more separated from the bulk layers and becomes more concentrated. More information about the electronic properties throughout the displacement can be found in reciprocal space. The band structure near the K-point for the six-layer graphite structure (without any top layer displacement) is shown in Fig. 5(r). As expected, all the bands are parabolic. As the top layer moves vertically, the band structure does not show any linear behavior at the Dirac point until Fig. 5(v) and (w). This marks the important electronic characteristic of graphene.

Note there is an extra set of linear bands coming from the odd number of layers remaining in the split-off graphite structure.[33] After analysis of the band structure throughout the movement of the top layer, we estimate that, around 0.090 nm, the unique electronic properties of graphene are fully present. Namely, the bands near the K-point are linear, and the total surface charge density has increased to nearly the level of isolated graphene.

**E. Vertical displacement total energy considerations**

In addition to getting information about the real-space and reciprocal-space properties, we also calculated the total energy of the system as a function of top-layer displacement, as shown in Fig. 6(a). The displacement is reported as a percentage of the equilibrium interplanar separation (0.334 nm), or the uni-axial strain $\varepsilon_{zz}$. The energy curve increases smoothly over the range sampled, and it transitions from positive to negative curvature near a strain of 13.5% (or a displacement of 0.045 nm). This inflection point is identified with an arrow. The calculated energy needed to fully separate the unit cell is found to be approximately 50 meV. From our earlier estimates we found that the STM tip expended 50 eV to lift the top layer by 30 nm.[22] Thus, we can now estimate that about 1,000 unit cells were separated during the lift. If the graphene was simply vertically lifted, a circular region with a radius of about 10 nm would be affected. Since this is similar to the height of the lifted graphene, we believe that a much larger area may slide across the graphite surface.

Next, we can estimate the force required to separate the layers by taking the derivative of the energy curve according to the Hellmann-Feynman theorem. This force (or uniaxial stress $\sigma_{zz}$) is a result of the attractive force between the graphitic layers, which increases up to the inflection point in the energy and subsequently decreases as shown in Fig. 6(b). The peak force required to separate the (1x1) layers is around 0.07 nN. This is smaller than the estimated electrostatic force applied by the STM tip (4 nN), which is consistent with the tip being able to lift the layer.

The charge densities found on the A atom site ($\rho_A$) and the B atom site ($\rho_B$) as a function of layer separation are presented in Fig. 6(c). These parameters have been normalized in two ways. First, since the total electronic charge in the top layer increased with the vertical displacement, every charge density was divided by the total charge density at that point, $\rho_{tot} = \rho_A + \rho_B$. This ensures that we track only changes in the relative charge densities ($\rho_A/\rho_{tot}$ and $\rho_B/\rho_{tot}$). Second, a normalization was applied to the data for each atom so that the normalized quantities, $N(\rho_A/\rho_{tot})$ and $N(\rho_B/\rho_{tot})$, vary from 0 to 0.5 and from 0.5 to 1, respectively. Thus, at zero displacement, $N(\rho_A/\rho_{tot})$ is a minimum, and $N(\rho_B/\rho_{tot})$ is a maximum, consistent with the STM images. Also, at the maximum displacement, the charge densities have equalized, also as seen in the STM images. (Note that these values are independent of the isovalue chosen for the simulated STM images.) A key benefit of this normalization scheme is that $N(\rho_B/\rho_{tot})$ represents a stepwise measurement of the decreasing interplanar coupling strength. If rescaled from 1 to 0, this parameter can be thought of as the effective mass scaling parameter.[34] The other parameter, $N(\rho_A/\rho_{tot})$, tracks the symmetry of the unit cell charge density. This parameter is tending toward zero as the symmetry between the A and B atoms is being broken. In this sense, this parameter (if rescaled from 0 to 1) represents the order parameter for the electronic reconstruction. The charge density profiles were also studied as a function of the bias voltage. For lower bias voltages (i.e., states closer to the Dirac point) the charge densities still began deviating from 50% at a strain around 40%, but the change to 1 or 0 happened more rapidly. This indicates that the states closer to the Fermi level are more sensitive to the surrounding environment.

The total charge density in each plane of atoms of the six-layer slab of graphite is shown in Fig. 6(d). The six sets of double peaks represent the $p_z$-orbitals around each atomic plane. The two end set of peaks are about half the size of the bulk peaks. Superimposed at the 0 nm position is the graphene charge density. Notice, the magnitude of the graphene charge density is more than three times larger than graphite at the surface. This is because, for graphite, the charge density at the surface is pulled into the bulk layers. Also, notice the charge density for graphene is significantly wider than that for graphite. Overall, the more massive charge density of graphene is also responsible for its high current carrying capacity and thermal conductivity. Two simulated STM images (shown in side view) for the surface of graphene and graphite are included as insets. The large charge density of graphene pushes its surface (imaged by the STM tip) further out into the vacuum.[29]

In a broader context, we are modeling the case where a normal force is continuously applied to the graphene as it approaches graphite. The two systems eventually begin to interact, and the graphene transitions to a layer of graphite. Interestingly, if pressure were applied still further, a second transition would occur from graphite to diamond,[35] as has been recently verified experimentally using femtosecond laser pulses to achieve the change.[36] However, what makes the graphene to graphite transition special is that it is the only known system where one can observe with atomic resolution how the electron acquires mass; or alternatively, how the electron loses mass and graphene generates the giant charge density responsible for its high current carrying capacity and thermal conductivity.

## IV. HORIZONTAL SHIFTING OF TOP GRAPHITE LAYER: BERNAL AND RHOMBOHEDRAL STACKING

So far, we have presented the EM-STM method and discussed the top graphite layer being lifted *vertically* above the bulk. However, by scanning the STM tip, it is possible to horizontally shift the top layer,[37] also resulting in a graphene-like electron charge density on the surface of graphite. In this section, we systematically study horizontal shifting, which also causes the transition between different stacking arrangements on the graphite surface. DFT calculations are completed to generate simulated STM images and analyze the likelihood for each pathway.

**A. ABA, ABC, and ABB stacked layers: Experiment**

Evidence for horizontal shifting of the top graphite layer along the carbon-carbon bond axis is shown in Fig. 7. Five characteristic atomic-resolution STM images are shown in the leftmost column, and each shows a different symmetry. Anyone that has done STM on HOPG has observed at one time or another STM images that are not the typical outcome shown in Fig. 7(c). We have systematically cataloged all of these images into groups in order to see a pattern. For example, in certain areas of the surface we can scan the tip along the C-C bond axis direction, and the image changes from a triangular pattern to a row-like pattern as shown in Fig. 7(b). To ensure this was not due to tip asymmetry but was due to the sample, this same tip was repositioned to other areas of the sample (a few mm away), and the experiment was repeated. We found the effect was associated with that specific region of the sample and did not follow the tip. Observation of a graphene-like surface charge density on graphite is shown in Fig. 7(a). STM images showing unequal intensity between the two carbon sites along the scan direction are shown in Fig. 7(d) and 7(e).

**B. ABA, ABC, and ABB stacked layers: Theory**

To gain insight into the origin of these results, simulated STM images were again extracted from DFT calculations without modeling the STM tip. This time, instead of moving the top layer vertically away from the bulk, it was slid along the surface. Specifically, the top layer of graphite was shifted along the C-C bond axis in units of a fraction of the bond length. The shifting process is divided into six steps, shown in the form of fraction of a bond length [see the top left corner of Fig. 7(f-j)]. The top view of simulated STM images are shown in Fig. 7(f-j). With the top layer of the graphite in its equilibrium position, a simulated STM image was generated and is shown in Fig. 7(h), with a ball-and-stick model for the top two layers shown in Fig. 7(m). After the top layer of graphite is shifted by half a bond length to the left, the charge density of the B-type atom is still larger than that of the A-type atom, as shown in Fig. 7(i). The stacking pattern is shown with a ball-and-stick model in Fig. 7(n). In this pattern, the B-atom in the top layer has moved away from the center of the hexagon of the lower layer and closer to the atom underneath. This increases the interaction between them and decreases the charge density of the B-atom in the top plane. On the other hand, the A-atom has shifted farther from the atom underneath, resulting in a weaker coupling between them which increases its charge density. After the top layer is shifted by one full bond length along this same direction (left), the first two layers of graphite completely overlap, and all the carbon atoms in the top plane are interacting equally with all the carbon atoms of the lower plane [see the model shown in Fig. 7(o)]. A honeycomb structure is displayed in the simulated STM image in Fig. 7(j). Notice that the electron charge density of the A-atom is slightly different from the B-atom. This is because of a subtle effect coming from the third layer. One pair of overlapping atoms is sitting directly above a carbon atom in the third layer, while the other pair of overlapping atoms is sitting directly above the hole at the center of the hexagon in the third layer. Excellent agreement is again found between the experimental STM data and the simulated STM data.

Next, in the computer model, the top layer of the graphite was shifted 0.30 bond lengths along the C-C bond-axis to the right this time, and another STM image was simulated as shown in Fig. 7(g). A row-like structure similar to Fig. 7(b) can be seen. The stacking arrangement for this configuration is best observed in the model shown in Fig. 7(l). This row-like structure is attributed to the overlap between the carbon atoms in the different layers. For a shift of 0.50 bond lengths, the charge density around all sites has equalized, as seen in Fig. 7(f). This gives the nice honeycomb structure and has excellent agreement with the STM data shown in Fig. 7(a). For the ball-and-stick model shown in Fig. 7(k), the top-layer carbon atoms are now exactly centered over the benzene rings in the layer below. We define this symmetric configuration as the "no overlap" structure since none of the carbon $p_z$ orbitals overlap with the ones in the plane below. Note, even though the "AA" stacking also yields a graphene-like surface charge density, we believe that the horizontal shift is not in this direction but is toward the "no-overlap" direction. Band structure near the K-point is shown as a function of horizontal displacement in Fig. 7(p-t). The most interesting results can be seen in the first and last diagrams. Notice that the typical linear dispersion associated with graphene is not present in either case, even though the simulated real-space STM images show the hexagonal symmetry of graphene. This is because the top layer is still bonding with the substrate, and a significant amount of the top layer surface charge density is still spread out into the bulk.

**C. Pathway from ABA to ABC stacked layers**

Evidence that the pathway between ABA and ABC is happening along the C-C bond-axis direction and specifically toward the "no overlap" structure is presented in Fig. 8. A key STM image of the HOPG surface is shown in Fig. 8(a). This image has "caught" the sudden transition from a graphite-like surface charge density to a graphene-like surface charge density. On the left side of the image is the signature triangular lattice of graphite, while on the right side is the distinct honeycomb structure of graphene. As mentioned earlier, previous work has demonstrated graphene formation on graphite; however, it was manifested as a continuous transition across the image. Our image is the result of a sudden jump, with the slow scan direction oriented horizontally now (left to right), in which a discrete change has occurred about two-thirds of the way through. Note the previous scan in this location showed all graphite while the subsequent scan showed all graphene. From this data we can not only measure the vertical shift in the top layer position, but we can also measure a horizontal shift as discussed next.

The energy per carbon atom found from DFT while shifting the top layer in ten equally spaced steps from ABA to ABC is shown in Fig. 8(b). Note, at each step the top layer was allowed to relax perpendicular to the surface to find the lowest energy pathway. The top layer is shifted along the C-C bond axis direction but directly toward the "no overlap" configuration. For this shift the ABC arrangement occurs after one bond length, and the "no overlap" situation is at the halfway point. Notice the symmetric shape of the energy curve indicates that there is no difference between the top surface layer shifting from ABA toward "no overlap" and from ABC toward "no overlap." However, when the top layer is shifted in the opposite direction, starting with the ABA stack, the DFT energy per atom is very different as shown in Fig. 8(c). When shifting in this direction, the top layer must move 2 bond lengths before reaching ABC stacking and the halfway point is the well-known AA stacking configuration. Notice, the barrier height for the "no overlap" direction is about 1.3 meV/atom in Fig. 8(b), while the opposite direction yields a barrier height of about 17 meV/atom in Fig. 8(c). Ball-and-stick models illustrating the stacking arrangement for the ABA, halfway point, and ABC configurations are also shown. Halfway through the shift shown in Fig. 8(b) is the point where the "no overlap" structure occurs between the top two layers [as more clearly illustrated in Fig. 7(k)] and a graphene-like surface charge density exists. Notice that the energy curve has a relatively flat top at this point, which may result in a meta-stable state, allowing the STM tip to occasionally image this higher-energy configuration. In fact, we believe the electrostatic attraction to the STM tip helps stabilize this configuration.

Overlaid on top of the inset image [Fig. 8(b)] are two ball-and-stick models of the graphene structure. The left-hand model is fit to the graphite image, while the right-hand model is fit to the graphene image. The overlapping region of the two models shows the best-fit horizontal displacement that has occurred during the movement of the top layer. This shows that the top layer of the graphite was, in fact, shifted in the low energy direction previously discussed and by an amount that corresponds to half a bond length.

By analyzing the coordination number of the top layer atoms with the second layer atoms, one can understand why the "no overlap" configuration should have the lowest energy graphene-like surface charge density. In both the ABA and ABC stacks, half the atoms have 6 nearest neighbors (with the plane below) and half the atoms have only 1, for an average coordination number of 3.5. When in the "no overlap" configuration, the atoms each have 3 nearest neighbors, for an average coordination of 3. When in the AA stacking configuration, the atoms each have 1 nearest neighbor, for an average coordination of 1. Since the atoms prefer to have a higher coordination number, the "no overlap" configuration energetically prefer over the AA stacking structure.

## V. CONCLUSION

We have shown that EM-STM measurements can be used to reversibly and irreversibly alter a graphite surface with considerable precision by varying the STM tip bias relative to the grounded sample. This technique was employed to physically alter the graphite surface with precise spatial control. In addition, it was used to controllably lift the top graphite layer away from the bulk. DFT simulated STM images for various displacements of the top layer relative to the bulk gave excellent agreement with experimental STM images. Band structure information predicted that the electronic properties of the top layer matched graphene after a vertical displacement of 0.090 nm. Finally, by using the theoretical real-space charge densities to characterize the transition from graphite to graphene, a stepwise model of the interplanar coupling that is responsible for the electron acquiring mass was presented. We also presented experimental evidence for a horizontal shift in the top layer of a graphite sample, induced by the STM tip and resulting in the observation of a graphene-like surface charge density. This shift is direction dependent and in excellent agreement with first-principles DFT energy calculations and simulated STM images. Specifically, the lowest-energy barrier direction is toward "no overlap" of the $p_z$ orbitals, while the highest-energy barrier direction is toward AA stacking (full overlap of the $p_z$ orbitals).

## ACKNOWLEDGEMENTS

P.X. and P.T. gratefully acknowledge the financial support of the Office of Naval Research (ONR) under grant number N00014-10-1-0181 and the National Science Foundation (NSF) under grant number DMR-0855358. Y.Y. and L.B. thank ARO Grant W911NF-12-1-0085, the Office of Basic Energy Sciences, under contract CR-46612 for personnel support and NSF grants DMR-0701558 and DMR-1066158; and ONR Grants: N00014-11-1-0384 and N00014-08-1-0915 for discussions with scientists sponsored by these grants. Calculations were made possible thanks to the MRI grant 0959124 from NSF, N00014-07-1-0825 (DURIP) from ONR, and a Challenge grant from HPCMO of the U.S. Department of Defense.

**Figure Captions**

FIG. 1. Constant current STM images of graphite acquired with a tip bias of +0.1 V. (a) 100 nm × 100 nm image showing a monolayer step separating two flat terraces. (b-f) 6 nm × 6 nm STM images showing: (b) triangular symmetry, (c) hexagonal symmetry, (d) starting triangular and ending hexagonal, (e) line-by-line tilted surface with triangular on left and hexagonal on right, and (f) starting hexagonal and ending triangular.

FIG. 2. (a) EM-STM measurement on HOPG showing STM tip height as a function of the tip bias. Measured tunneling current is shown as an inset. (b-d) Schematic illustrations showing the STM tip lifting the surface layer of a graphite sample as the bias voltage increases from (b) 0.2 V to (c) 0.6 V to (d) 0.8 V.

FIG. 3. (a-f) Chronological series of about 150 nm × 150 nm filled-state STM images of one location on graphite surface taken with a tip bias voltage of +0.1 V and a setpoint current of 1.0 nA. EM-STM measurements were performed on the white strip during the acquisition of the images shown in (b) and (d). (g) EM-STM measurement on the white strip showing STM tip height as a function of tip bias. Measured tunneling current is shown as an inset. (h) Calculated electrostatic force exerted by the STM tip on the graphite surface as a function of the tip height. Area under curve corresponds to the total energy expended by the STM.

FIG. 4. Five characteristic EM-STM measurements on HOPG taken at various setpoint currents: (a-b) 0.1 nA, (c-d) 0.2 nA, and (e) 1 nA. (f-j) Calculated electrostatic force shown as a function of STM tip height for each EM-STM data set (a-e), respectively.

FIG. 5. (a-e) Filled-state atomic-resolution STM images of the HOPG surface taken with a tip bias voltage of 0.1 V and setpoint current of 0.2 nA. Notice that (a) shows triangular symmetry, while (e) shows hexagonal symmetry. (f-k) Simulated STM images of graphite taken from DFT calculations. The top layer's vertical displacement from the equilibrium position is indicated at the top left corner. (l-q) Side views of the simulated six-layer graphite structure shown with the top layer separated from the bulk by different amounts. (r-w) Band structure near the K-point for the six-layer graphite structure as the top layer separates from the bulk. Notice the bottom diagram has a linear band structure characteristic of graphene.

FIG. 6. (a) DFT energy per unit cell as a function of vertical displacement of the top layer of graphite. (b) Force as a function of displacement obtained by taking the derivative of the energy with respect to displacement. (c) Normalized charge density on the A site and B site, taken from DFT simulated images, as a function of displacement. (d) Total charge density as a function of lattice position for both graphite (6 layer slab) and graphene (positioned at 0 nm). Two side-view simulated STM images are shown to scale as insets. Notice the top layer of graphite is much thinner than graphene.

FIG. 7. (a-e) Filled-state atomic-resolution STM images of HOPG surface taken with bias voltage of 0.1 V and setpoint current of 0.2 nA showing different symmetries; (a) hexagonal, (b) row-like, (c) triangular, (d) brick-wall like, and (e) hexagonal with slight asymmetry. (f-j) Simulated STM images of graphite taken from DFT calculations. The top graphite layer is horizontally shifted along the C-C bond direction. The net displacement is indicated in the top

left corner in fractions of a bond length. (k-o) Ball-and-stick structural models showing the shift of the top graphite layer relative to the second layer along the C-C bond direction indicated by the arrow. (k) Shows the "no overlap" structure, (m) shows the normal ABA graphite stack, and (o) shows the "AA" stacking pattern. (p-t) Band structures shown near the K-point for the simulated six-layer graphite structures, but with various top layer shifts indicated in (f-j), respectively.

FIG. 8. (a) STM image displayed with the slow scan direction going left to right. The STM image starts off showing a triangular pattern and about two-thirds through the image shows a hexagonal pattern. (b) DFT energy/atom is plotted (squares) as a function of the top layer of graphite horizontal shift in the direction illustrated by the three ball-and-stick models shown. The pink frame represents the top layer, the green frame represents the second layer, and the blue frame represents the third layer. Inset image: STM data showing graphite- and graphene-like surface charge densities in a single scan [cut from (a)]. Ball-and-stick models are overlaid on the graphite and graphene surfaces to indicate the locations of the carbon atoms as well as the direction and magnitude of the horizontal shift that occurred (the white box behind the models highlights the shift between them). (c) DFT energy/atom is plotted (circles) as a function of the top layer horizontal shift in the direction indicated by the three ball-and-stick models shown.

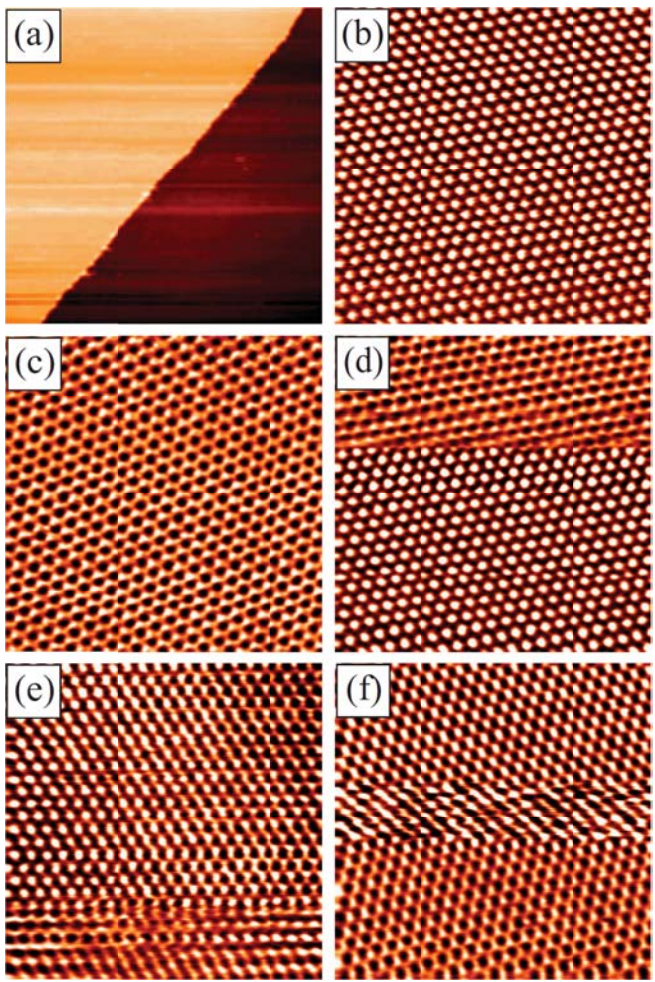

FIG. 1

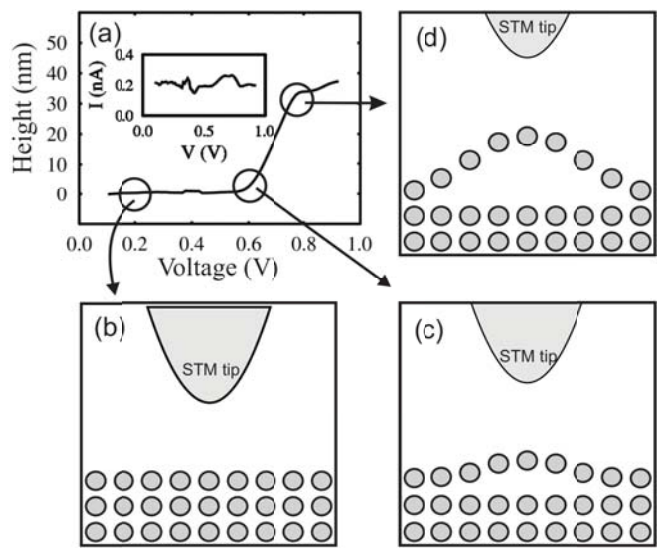

FIG. 2

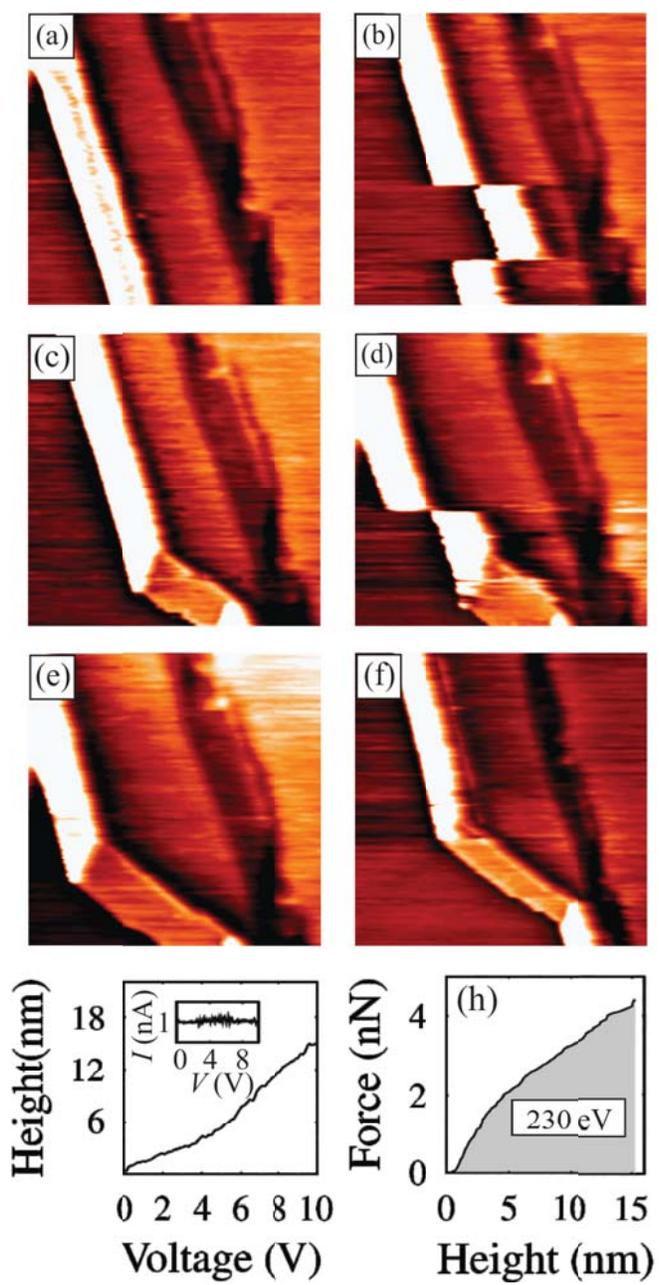

FIG. 3

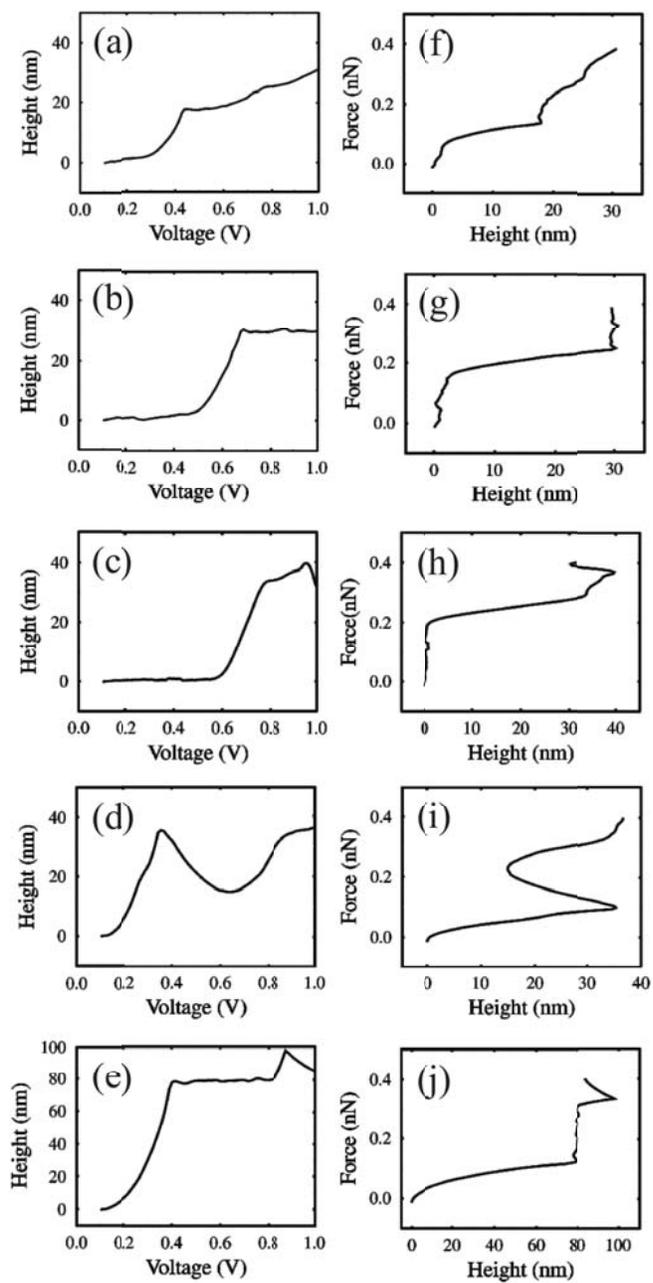

FIG. 4

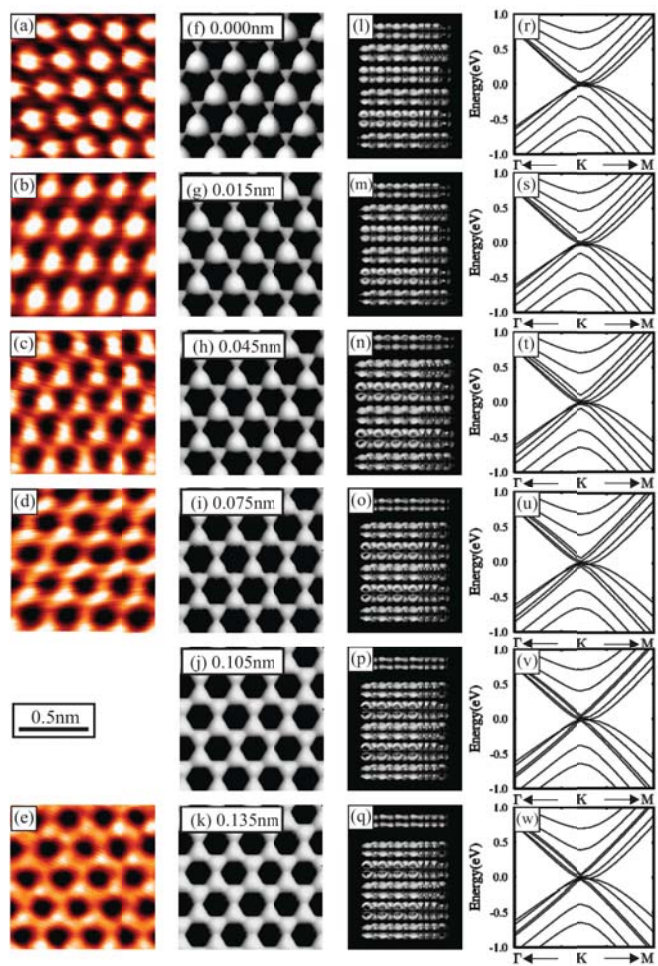

FIG. 5

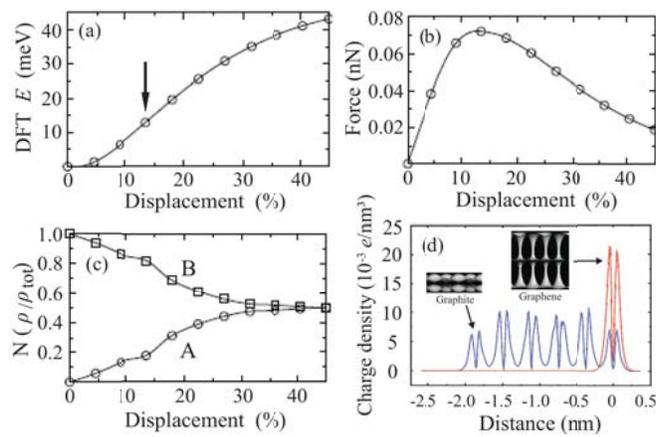

FIG. 6

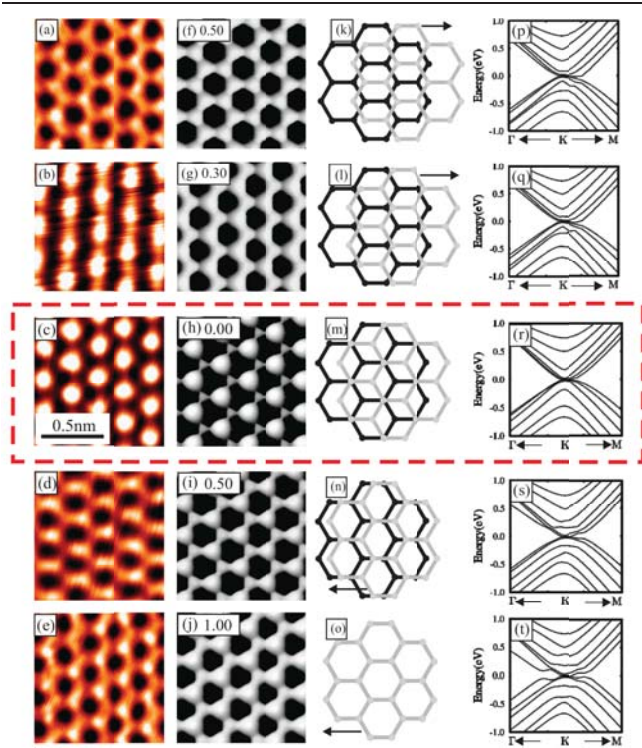

FIG. 7

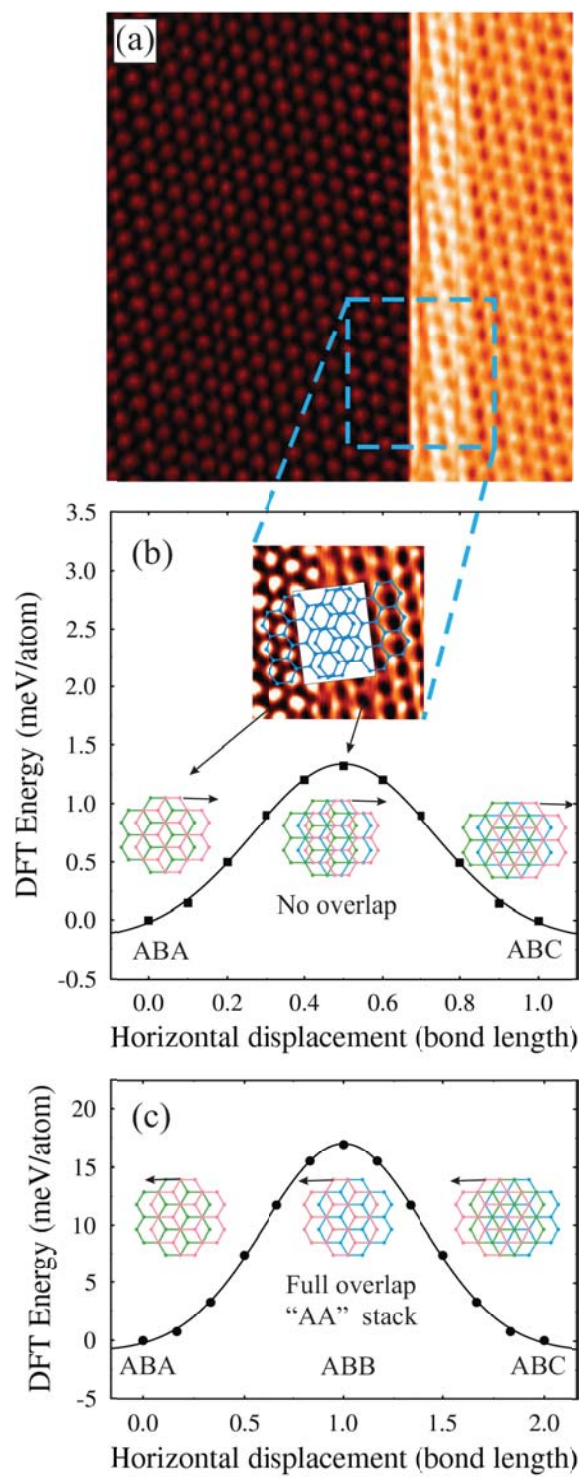

FIG. 8